\documentclass[conference]{IEEEtran}
\IEEEoverridecommandlockouts
\usepackage{cite}
\usepackage{amsmath,amssymb,amsfonts}
\usepackage{algorithmic}
\usepackage{graphicx}
\usepackage{textcomp}
\usepackage{xcolor}
\usepackage[a4paper, total={184mm,239mm}]{geometry}
\def\BibTeX{{\rm B\kern-.05em{\sc i\kern-.025em b}\kern-.08em
    T\kern-.1667em\lower.7ex\hbox{E}\kern-.125emX}}
\begin{document}

\title{Extended Abstract: Synthesizable Low-overhead Circuit-level Countermeasures and Pro-Active Detection Techniques for Power and EM SCA \\
}

\author{\IEEEauthorblockN{Archisman Ghosh}
\IEEEauthorblockA{\textit{ECE, Purdue University, West Lafayyete, IN-47906, USA} \\
email: ghosh69@purdue.edu, phone: +17657758963 \\
Promoter: Shreyas Sen
}
}
\maketitle

\section{Introduction}
The gamut of today’s internet-connected embedded devices has led to increased concerns regarding the security and confidentiality of data. Most internet-connected embedded devices employ mathematically secure cryptographic algorithms to address security vulnerabilities. Despite such mathematical guarantees, as these algorithms are often implemented in silicon, they leak critical information in terms of power consumption, electromagnetic (EM) radiation, timing, cache hits and misses, photonic emission and so on, leading to side-channel analysis (SCA) attacks. This thesis focuses on low overhead generic circuit-level yet synthesizable countermeasures against power and EM SCA. Existing countermeasures (including proposed) still have relatively high overhead which bars them from being used in energy-constraint IoT devices. We propose a zero-overhead integrated inductive sensor which is able to detect i) EM SCA ii) Clock glitch-based Fault Injection Attack (FIA), and iii) Voltage-glitch based Fault Injection Attack by using a simple ML algorithm. Advent of quantum computer research will open new possibilities for theoretical attacks against existing cryptographic protocols. National Institute of Standard \& Technology (NIST) has standardized post-quantum cryptographic algorithms to secure crypto-systems against quantum adversary. I contribute to the standardization procedure by introducing the first silicon-verified Saber (a NIST finalist modulo Learning with Rounding scheme) which consumes lowest energy and area till date amongst all the candidates. 

\section{Power \& EM SCA security by Digital Signature Attenuation}
\begin{figure}[!t]
  \centering
   \includegraphics[width=0.5\textwidth]{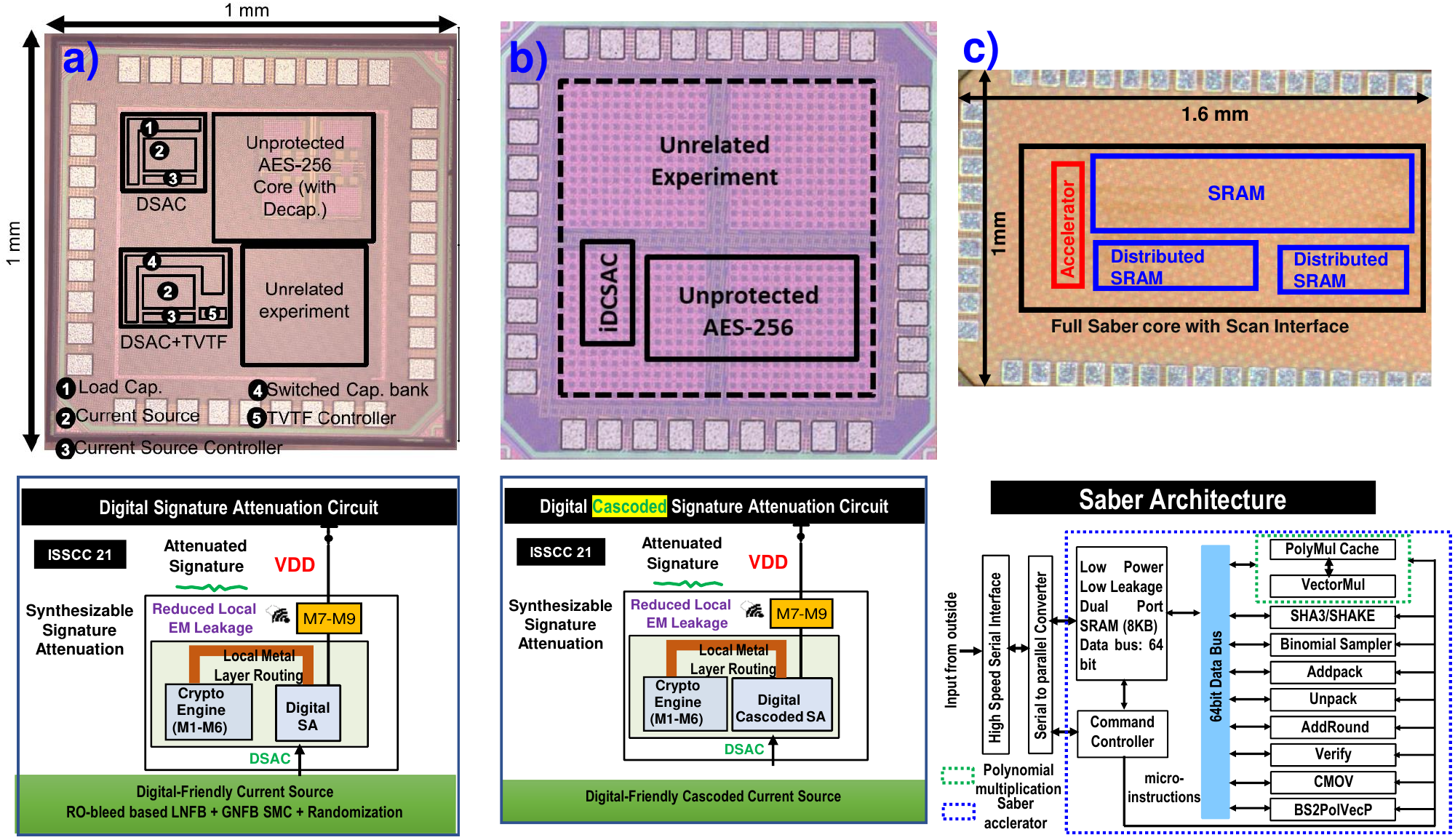}
   \caption{{a) 65nm CMOS IC for digital signature attenuation technique and single on-chip sensor for EM SCA \& FIA detection. b) 65nm CMOS IC for sustainable digital signature attenuation circuit with intelligent attack detector. c) 1st Silicon verified Saber IC and architecture.}}
   \label{chip_micro_dia}
 \end{figure}

Circuit-level countermeasures against power/EM SCA include current equalizer, series Low-Dropout Regulator, Integrated Voltage Regulator, enhancing protection up to 10M minimum traces to disclosure (MTD). Recently, current domain signature attenuation and randomized NL-LDO cascaded with arithmetic countermeasures achieved $>1B$ MTD with a single and two countermeasures, respectively. This work for the first time brings the benefit of signature attenuation in the digital domain and extends the state-of-the-art of circuit-level countermeasure by 25\% by providing $>1.25B$ MTD for the first time and 25x with respect to single digital countermeasure (250M MTD) \cite{ghosh202136, ghosh2021syn} (Attached paper). Digital Signature Attenuation circuit (DSAC) utilizes a synthesizable Current Source (CS) which is the key component to attenuate signature to achieve high MTD (20M MTD standalone). Multiple parallel multi-stage ring oscillators (ROs) are used as the bleed path which assist in 1) stabilizing the internal AES node voltage, 2) Global Negative Feedback (GNFB) fully digital Switched Mode Control (SMC) Loop which stabilizes the number of CS slices in case of sudden change in power consumption and 3) RO-bleed strength randomization which adds up security through power supply noise injection (250M MTD when added to signature attenuation). Synthesizable CS slices are made of stacked power gates which are biased by self-connected inverter. An additional technique of time domain obfuscation at VDD port, namely Time Varying Transfer Function (TVTF) \cite{tcas2switchcap, ojcas_switch_cap} is used for further security ($>1.25B$ MTD).  Moreover, the encryption engine is routed through local metal layer (upto Metal 6 in 65nm) to reduce meaningful EM emanation which helps in EM SCA resiliency. The entire solution is demonstrated in a 65nm test IC (Fig.~\ref{chip_micro_dia}(a)) for real life applications. 
\section{Self-Awareness: EMSCA \& FIA Detection by single on-chip loop \& ML algorithm}
Though, the previous solution provides a very high experimental physical security against side channel analysis and extremely low area \& power overhead (25\% for DSAC, 52\% for DSAC+TVTF), energy overhead can still be reduced through a successful detection of SCA and FIA. In the next work \cite{ghosh2022sca,ghosh2022electromagnetic}, we develop a pro-active approach to detect and counter these attacks by embedding a single on-chip integrated loop around a crypto core (AES-256), designed, and fabricated using TSMC 65nm process shown in FIg.~\ref{chip_micro_dia}(a). The measured results demonstrate that the proposed system has three capabilities: 1) it provides EM-Self-awareness by acting as an on-chip H-field sensor, detecting voltage/clock glitching fault-attacks; 2) it senses an approaching EM probe to detect any incoming threat; and 3) it can be used to induce EM noise to increase resilience against EM attacks. This work combines EM analysis, ML based secured system and shows the efficacy by measurements from custom-built 65nm CMOS IC. A simple FCN based ML algorithm can detect FIA or EM SCA in this framework with nearly 100\% accuracy once being trained with approximately 3000 traces from the on-chip sensor. 
\section{Sustainable Signature Attenuation Technique with intelligent voltage drop attack detector}
DSAC shows the promises of generic low-overhead and synthesizable countermeasure with a synthesizable current source as source of signature attenuation. The next work demonstrates an analog-like digital cascoded signature attenuation technique for further attenuation in digital domain to achieve 200M MTD just by signature attenuation. Also, we explore the possibility of practical attack on signature attenuation-based countermeasure.  High suppression is achieved by the current sources (CS) in saturation in this category of countermeasures. For an intelligent attacker, supplying the same amount of current while maintaining the CS in linear region will reduce the signature attenuation and increase the correlated power/EM leakage. If an attacker reduces VDD from the desired value such that the CS goes to the linear region while the lower threshold and upper threshold of the global SMC loop remain constant, more CS slices will be turned on due to the SMC operation, drawing the same amount of current (unlike CS being in saturation) from the supply pin, to operate AES-256. At that optimal region for attack, the CS will remain in linear region with the AES operating correctly, thus enabling an intelligent attacker to reduce the efficacy of signature attenuation and break the secret key with a low number of traces (105K MTD, instead of ~200M). However, lowering VDD by a significant amount will cause the SMC loop failing to supply the required current. Hence, AES voltage will drop continuously. It will not be able to work failing the attack. Hence, attackers need to find the attack point precisely. This work also provides a solution to this type of intelligent voltage drop attack. An intelligent attack detector monitors the voltage (divided to approximately meet voltage at AES local power node) at VDD node using a RO. Another RO based voltage tracker tracks the voltage at AES node.  A sudden drop of voltage causes a mismatch in the comparison circuit and rings an alarm about the possibility of such an attack. An experiment on 65nm CMOS IC (Fig.~\ref{chip_micro_dia}(b)) \cite{ghosh2022digital, ojsscs_rstellar} shows such an attack can be detected within 0.8ms making CPA improbable. 
\section{1st Silicon Verified ASIC of Saber with Striding Toom-Cook multiplication with Lazy Interpolation}
Progress in quantum computing research enlightens that arrival of large-scale quantum computers will break the security assurances of practically deployed public-key cryptography. Lattice based cryptography is a brunch of public key cryptography, of which the mathematical investigation (so far) resists attacks with quantum computers. By choosing a module learning with errors (MLWE) algorithm as the next standard, NIST follows this approach. The multiplication of polynomials is the central bottleneck in the computation of lattice-based cryptography. Because one popular use of public key cryptography is to establish common secret keys, the focus of this work is on compact area, power and energy budget. While most other work focuses on optimizing number theoretic transform (NTT) based multiplications, in this thesis we highly optimize a Toom-Cook based multiplier. Here, a memory-efficient striding Toom-Cook with lazy interpolation is demonstrated, results in a highly compact, low power implementation, which in addition enables a very regular memory access scheme. To demonstrate the efficiency, we integrate this multiplier into a Saber post-quantum accelerator (1st Silicon proven Saber ASIC (Fig.~\ref{chip_micro_dia}(c)) \cite{ghosh2022334uw, ghosh2023334}), one of the four NIST standardization procedure finalists. Algorithmic innovations to reduce active memory, timely clock gating and shift-add multiplier has helped to achieve 38\% less power than state-of-the art PQC core, 4 $\times$ less memory, 36.8\% reduction in multiplier energy and 118$\times$ reduction in active power with respect to state-of-the-art Saber accelerator (not silicon verified). This accelerator consumes $0.158mm^2$ active area which is lowest reported till date despite process disadvantages of the state-of-the-art designs.
\section{Acknowledgment \& Notes}
I am grateful to my promoter Prof. Shreyas Sen for giving me the opportunity to compete in a different PhD forum. This extended abstract is archived for educational purposes as an example for different PhD forum competitions. Key intuitions behind different techniques can be found in \cite{sscmagazine} and details of the security techniques can be found in our conferences, journals as well as in my thesis \cite{Ghosh2024}.

\bibliographystyle{unsrt} {
    \bibliography{phd_forum.bib}
}

\end{document}